\documentclass[arxiv]{nesy2025} 

\usepackage{longtable}

\usepackage{booktabs}

\usepackage{algorithm}
\usepackage{algpseudocode}

\usepackage{booktabs}
\usepackage{multirow}
\usepackage{multicol}

\algdef{SE}[VARIABLES]{Variables}{EndVariables}
   {\algorithmicvariables}
   {\algorithmicend\ \algorithmicvariables}
\algnewcommand{\algorithmicvariables}{\textbf{global variables}}

\algblock{Input}{EndInput}
\algnotext{EndInput}
\algblock{Output}{EndOutput}
\algnotext{EndOutput}
\newcommand{\Desc}[2]{\State \makebox[6em][l]{#1}#2}
\newcommand{\explanation}{\mathcal{E}}

\usepackage{adjustbox}
\usepackage{subcaption}

\usepackage{newfloat}
\usepackage{listings}
\DeclareCaptionStyle{ruled}{labelfont=normalfont,labelsep=colon,strut=off} 
\floatstyle{ruled}
\newfloat{listing}{tb}{lst}{}
\floatname{listing}{Listing}

\usepackage{xcolor}
\usepackage[load-configurations=version-1]{siunitx} 

\theorembodyfont{\upshape}
\theoremheaderfont{\scshape}
\theorempostheader{:}
\theoremsep{\newline}

\title[MetaExplainer: A Framework to Generate Multi-Type User-Centered Explanations]{MetaExplainer: A Framework to Generate Multi-Type User-Centered Explanations for AI Systems}
\author{\Name{Shruthi Chari}\Email{charishruthi@gmail.com} \addr Rensselaer Polytechnic Institute, Troy, NY \\
\Name{Oshani Seneviratne}  \addr Rensselaer Polytechnic Institute, Troy, NY \\
\Name{Prithwish Chakraborty} \addr Amazon Science, NY, NY \\
\Name{Pablo Meyer} \addr IBM Research, Yorktown Heights, NY \\
\Name{Deborah L. McGuinness} \addr Rensselaer Polytechnic Institute, Troy, NY}

\begin{document}

\maketitle
\thispagestyle{empty}

\begin{abstract}
Explanations are crucial for building trustworthy AI systems, but a gap often exists between the explanations provided by models and those needed by users. Previous research and our interactions with clinicians have shown that users prefer question-driven and diverse explanations. To address this gap, we introduce MetaExplainer, a neuro-symbolic framework designed to generate user-centered explanations. Our approach employs a three-stage process: first, we decompose user questions into machine-readable formats using state-of-the-art large language models (LLM); second, we delegate the task of generating system recommendations to model explainer methods; and finally, we synthesize natural language explanations that summarize the explainer outputs. Throughout this process, we utilize an Explanation Ontology to guide the language models and explainer methods. By leveraging advanced language models and a structured approach to explanation generation, MetaExplainer aims to enhance the interpretability and trustworthiness of AI systems across various applications, providing users with tailored, question-driven explanations that better meet their needs. 
Comprehensive evaluations of MetaExplainer demonstrate a step towards evaluating and utilizing current state-of-the-art explanation frameworks. Our results show high performance across all stages, with a $59.06\%$ F1-score in question reframing, $70\%$ faithfulness in model explanations, and $67\%$ context-utilization in natural language synthesis. User studies corroborate these findings, highlighting the creativity and comprehensiveness of generated explanations. Tested on the Diabetes (PIMA Indian) tabular dataset, MetaExplainer supports diverse explanation types, including Contrastive, Counterfactual, Rationale, Case-Based, and Data explanations. The framework's versatility and traceability from using ontology to guide LLMs suggest broad applicability beyond the tested scenarios, positioning MetaExplainer as a promising tool for enhancing AI explainability across various domains.

MetaExplainer codebase: \url{https://github.com/tetherless-world/metaexplainer}
\end{abstract}


\section{Introduction}
\label{sec:intro}
Explainable AI has been a cornerstone of research for decades and has evolved with AI approaches. However, with the increase in complexity of AI methods, the results of explainer methods alone have been found insufficient for end-user needs. Several researchers~\cite{tonekaboni2019clinicians} have found that users require explanations for different purposes and as answers to various question types to effectively use explanations in their decision-making when interacting with AI systems. Further, several taxonomies~\cite{wang2019designing},~\cite{liao2020questioning} have been proposed to catalog the various explanation types and fewer to link these explanation types to AI methods that generate them. However, there remains a gap in generating explanations in real-time along various user-centered explanation types for user questions. Further, the existing conversational explanation approaches build on something other than the most well-cited model explainer outputs but are implemented for specific use cases, making them less generalizable.  



\begin{figure*}[hbt!]
    \centering
    \includegraphics[width=1.0\linewidth]{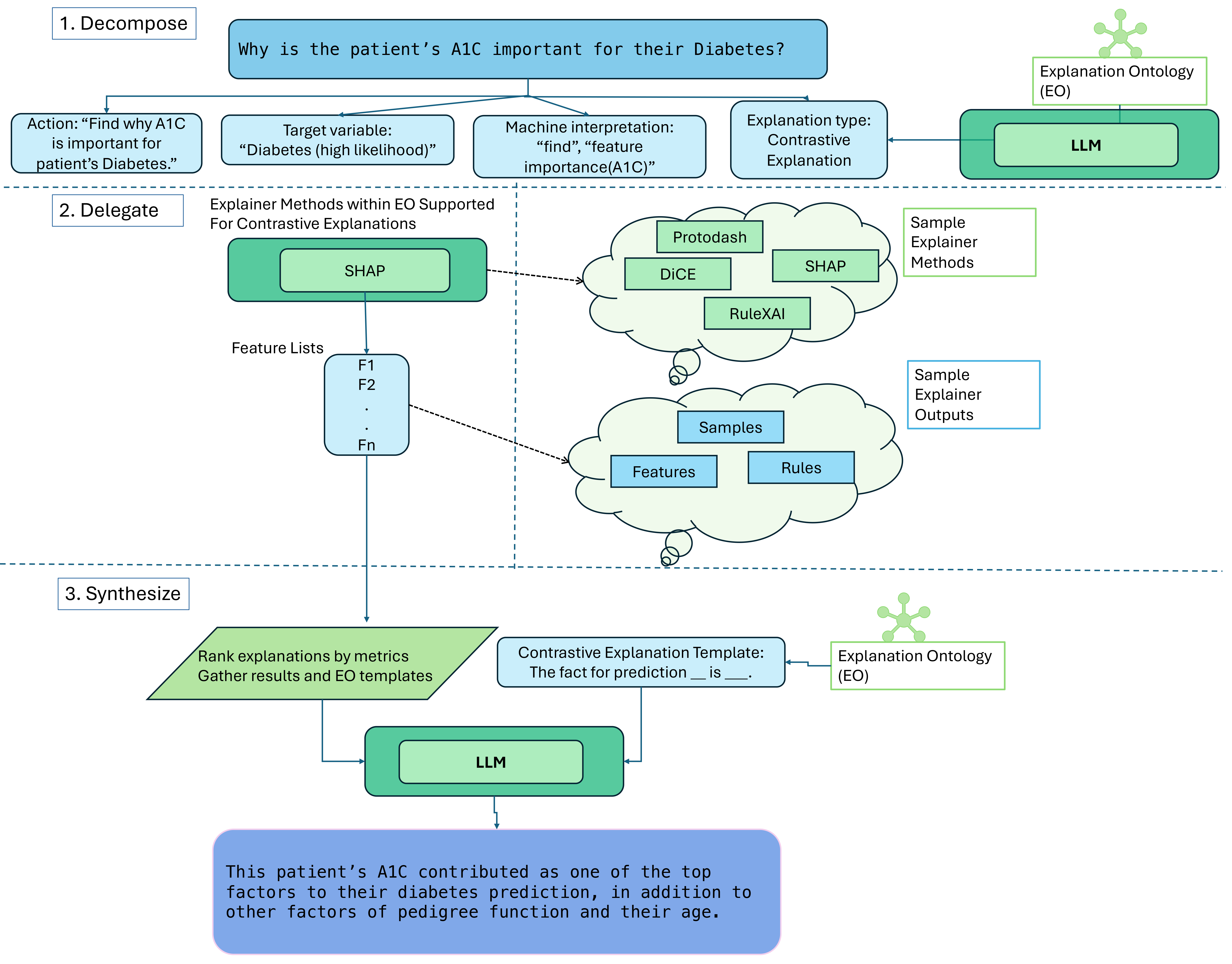}
    \caption{Workflow diagram of MetaExplainer, highlighting the different input and output streams. Methods are indicated by green boxes and all other data components are shown in blue.}
    \label{fig:MetaExplainer-workflow}
\end{figure*}

Hence, in the pursuit of supporting natural-language~\cite{lakkaraju2022rethinking}, diverse~\cite{mittelstadt2019explaining}, ~\cite{miller2019explanation} and multi-type user-centered~\cite{dey2022human} explainability, we present 
the development of the MetaExplainer (Fig. \ref{fig:MetaExplainer-workflow}), a multi-stage (decompose-delegate-synthesis) general-purpose explanation framework capable of generating explanations for end-user questions by summarizing explanations from several model explainers (such as SHAP~\cite{lundberg2017unified}, DiCE~\cite{mothilal2020explaining} as seen in Fig. \ref{fig:MetaExplainer-workflow}) in natural language. Specifically, we explore how user-driven reasoning and conversations can lead to satisfactory explanations of AI systems with the design and implementation of MetaExplainer. We address the following research issues: 
\begin{itemize}
    \item How to build an extensible explainability framework that provides appropriate and useful explanations drawing from different data modalities, knowledge sources, and explainer methods?
    \item How can the MetaExplainer be evaluated to support user-centered needs across representative use cases and metrics? 
\end{itemize}

In the rest of the paper, we first introduce relevant literature in explainable AI (Sec. 
\ref{sec:background}
) that motivated us to build MetaExplainer, describe the technical details of the MetaExplainer (Sec. \ref{sec:methods}), and present the results (Sec. \ref{sec:results}) of comprehensive experimental evaluations, both from a quantitative multi-stage point of view and 
qualitative evaluation from a small scale user-study.
We finally discuss (Sec. \ref{sec:discussion}) the benefits of a neuro-symbolic framework such as ours and takeaways from our results.

\section{Background} \label{sec:background}
\subsection{Model Explainers}
Model explainers or explainer methods provide model explanations of AI systems, and they typically explain model behavior on a per prediction level (LIME~\cite{ribeiro2016should}, SHAP~\cite{lundberg2017unified}, etc.) or globally about the model's functioning (Boolean Rule Column Generator~\cite{dash2018boolean}). However, there are several challenges with using these model explainers directly in use cases to support end-user explainability. The outputs of these model explainers have often been found to be insufficient or lacking in domain context for end-users such as domain experts to take action on~\cite{ghassemi2021false, tonekaboni2019clinicians}. Further, running the model explainers is better suited for system developers who understand the functioning of AI systems. Additionally, it is hard to decide on what model explainer to run and how to present the results to domain experts; hence, encoding or registering the details of model explainers in ontologies and presenting model explanation outputs in natural language via explanation types can be beneficial. Further, presenting model explanation outputs as natural-language~\cite{slack2023explaining} can help adapt explanations to different user groups within the same domain~\cite{dey2022human} (e.g., clinical researchers prefer data explanations over clinicians who prefer scientific explanations). In the MetaExplainer, we include a well-cited set of explainers (Tab. \ref{tab:metrics_summary}), with provisions and directions on how to support more. 

\subsection{Explanation Ontology}
In the Explanation Ontology (EO)~\cite{chari2023explanation}, we model the system-, user- and interface- dependencies of explanations of AI systems. The EO model helps represent fifteen user-centered explanations, which were introduced when explainability was identified as an essential prong of trustworthy AI. During this time, several researchers studied end-users and surveyed literature, finding that explanations serve different purposes for end-users and address various questions~\cite{wachter2017counterfactual},  ~\cite{miller2019explanation}, ~\cite{doshi2017accountability}. Hence, suggesting that there is an inherent diversity in the types of explanations. However, the descriptions of various explanation types were mainly contained in several papers and were difficult to support without a unified representation such as the EO. The EO addresses the gap by serving as a resource to represent explanations such that they can be constructed from the dependencies. There have been continuous updates to the EO across versions (current version being three\footnote{EO: \url{https://purl.org/heals/eo/3.0.0}}), to support model explainers and their outputs such that the explanations can be built from explainer methods that generate them. 
In the MetaExplainer, we explore how the EO can serve multiple purposes to support the generation of natural-language explanations, including identifying what explanation type would best address a user question, what explainer methods are capable of providing explanations for the identified type, and finally, what templates are suggested to structure explanations of the identified type. 

\subsection{Related Work} \label{sec:related_work}
Our closest related work is an effort from ~\cite{slack2023explaining},  MegaExplainer, to produce explanations directly for user questions. This is a framework wherein the authors use utterances or cues to prompt large language models (LLMs) to create filters that can be used to serve as inputs for model explainers and thus generate model explanations in line with user questions. 
In the MegaExplainer, users can ask questions to a TalkToModel interface and are provided natural-language explanations of model explanations.
However, their framework is limited by the types of questions that can be addressed owing to the rule-based generation process of questions. Our MetaExplainer is both modular and flexible and is hence adaptable to more explanation types and methods (i.e., developers can add support for more explainer methods under the Delegate stage of the MetaExplainer). Further, we currently support five user-centered explanation types (Case-based, Contrastive, Counterfactual, Data, and Rationale - Tab. \ref{tab:explanation_types} in Appendix), which is more than the four explanation types (Contrastive, Counterfactual, Data, and System Performance) that Megaexplainer supports. We reused the design principles of Megaexplainer where possible.

 Additionally, some researchers such as Krishna et al.~\cite{krishna2023towards} are also looking at when it is wise to produce explanations and when not~\cite{ghassemi2021false}, they propose Robust Counterfactual Explanations under the Right to be Forgotten (ROCERF) framework. 
 While ROCERF is a start at analyzing whether explanations provide value, it is hard to apply the method to problems beyond the field of counterfactual explanations studied. In the MetaExplainer, we evaluate explanation outputs using well-studied explanation metrics~\cite{zhou2021evaluating}, hence, providing a notion of their validity and usefulness (more findings in Sec. \ref{sec:results}). Also, McGuinness and Silva~\cite{mcguinness2007explaining},~\cite{mcguinness2004explaining} have early work on composing explanations from components in task-based environments. However, their work predates the explainer methods that we use today but is still relevant in the various modules, such as dispatchers, constraint, and knowledge explainers contributing to explanations. We leverage their modular design while designing the MetaExplainer.

\section{Methods} \label{sec:methods}

We adopt a modular approach to implement the MetaExplainer in that we break down the design into three stages; \textit{decompose:} convert the user question into a machine interpretation that can then be \textit{delegated:} to explainer methods registered against explanation types in EO and finally, the explainer outputs are \textit{synthesized} into natural-language explanations presented to the end-users. We implement the framework as a Python package, such that each of the stages can call methods and classes from one another (Listing \ref{lst:MetaExplainer}), backed by file-based communication,
enabling the stages to build off of the intermediary stage outputs as seen in Fig. \ref{fig:MetaExplainer-workflow} (an end-end example is presented in the Appendix (Sec. \ref{apt:end-end}). 


\begin{algorithm} [hbt!]
    \caption{Pseudocode for MetaExplainer indicating the three stages - Decompose, Delegate, and Synthesis, and their inputs and outputs.}
    \label{lst:MetaExplainer}
    \begin{algorithmic}[l]
		\Require Explanation Ontology (EO), Data Store (DS) \newline
        \State Explanation type - explainer graph, $G^1 = \{(t_i, em_j),  \forall i \in N \text{ and } \forall j \in M\}$  \newline
        \State Data type - explainer graph, $G^2 = \{(d_k, em_j),  \forall k \in K \text{ and } \forall j \in M\}$  \newline
        
        \Input
          \Desc{$uq$}{User Question}
        \EndInput  \newline
          
          \State $rq$ = \texttt{Decompose}($uq$) 
          \State $eq$ = \texttt{Delegate}($rq$) 
          \State $E$ = \texttt{Synthesis}($eq$)  \newline

        \Output
          \Desc{$rq$}{list of questions reframed from $uq$}
          \Desc{$E$}{list of explanations $\{E_e\}$ that answer $uq$ where 
          $E_e = \{ {\explanation}_e, t_e, em_e, rq_e, uq_e\}$ } 
          \EndOutput
    \end{algorithmic}
\end{algorithm}
\vspace{-0.5em}
\subsection{Stages of MetaExplainer Framework}
\textit{\textbf{Decompose:}} In the Decompose stage, the objective is to generate a machine-actionable parse of a user question (s), including the explanation type that best addresses the question and the features being asked about. For example, in a question - ``Why is a 60-year-old woman with a BMI of 28 more likely to have Diabetes?'' we would want to identify that this question is best addressed by facts in contrastive explanations and has filters applied on features - age, BMI. From the question, we could also infer that the patient's likelihood of Diabetes is high. These attributes, if captured, help ensure that the explainer methods can reliably address the user question. 

Hence, in the Decompose stage, we set up a task to convert a user question $uq$ into a reframed question $rq$, which is a tuple capturing the $\{$question, explanation type, machine interpretation, action$\}$. We choose predicate logic as an intermediate format between natural language and machine language~\cite{ali2009selecting} for machine interpretation as it captures variables and values in a logical format. We can then process the predicate logic in the further stages of the MetaExplainer to run explainers on variables and their values. Within the Decompose stage, we prompt GPT-3.5 Turbo first to generate a question bank of question ($uq$) -reframed question ($rq$) training pairs for each explanation type on a per dataset basis and verify them before using these pairs to instruction fine-tune LLama models to generate parses for unseen $uq$ (Fig. \ref{fig:prompt2}). We fit the $uq$:$rq$ pairs into user: response style into instruction prompts~\cite{taori2023stanford} and use Low-Rank adaptations~\cite{hu2021lora} and Supervised Fine-tuning Trainers (SFTTrainer)~\cite{gunel2020supervised} to fine-tune the Llama models. 



\textit{\textbf{Delegate:}} In the Delegate stage, the objective is to execute relevant explainer methods to address a user question ($uq$). Here, we further break it down into parsing tasks, explainers ($em$) execution tasks, and explainer outputs ($eq$) processing tasks. In the parsing task, we parse the machine interpretation field of the $rq$ for filters on features, the explanation type ($et$) field to identify what explainer methods ($em$) can be run, and the action field to identify if there are other actions the question asks about (e.g., preprocessing, accuracies, etc.) In the execution task, we run open-source explainer methods ($em$) (Tab. \ref{tab:metrics_summary}), some of which are available as part of the IBM AIX 360 toolkit~\cite{arya2022ai} (Protodash and SHAP). We leverage the explanation type - explainer method mappings in the EO to identify what explainers to run. For example, EO stores that contrastive explanations are on local explanation outputs from saliency methods~\cite{arya2019one}), and we extract these mappings via SPARQL queries triggered from within the MetaExplainer. Further, these explainers are typically run on trained models and their learned weights or predictions; hence, we also identify the best-performing model and run the explainers on these models (Sec. \ref{sec:results}). Finally, in the output processing step, we persist the explainer outputs as dataframes that the Synthesis stage can then read to include in natural-language explanations of the explanation type. Interestingly, we also find that the mappings between explainer method output $eq$ and modalities are non-unique (Tab. \ref{tab:metrics_summary}), and these modalities are useful to identify metrics to evaluate the explainers $em$ on as well as to define how these outputs can be translated to natural-language in the Synthesis stage.


\textit{\textbf{Synthesis:}} In the Synthesis stage, we aim to combine and synthesize the explainer output (s) into natural-language explanations aligned with the identified explanation type's ($et$) expected templates. For example, a contrastive explanation should contain facts, foils, or both supporting model predictions. However, the explainer outputs from Delegate are in dataframes and not in natural-language; and need to be converted into natural-language. This task can be broken down into three steps, including a retrieval step where we extract relevant data points from the explainer outputs to include in the final explanation, an augmentation/alignment step where we align the outputs to fit the identified explanation type's ($et$) templates, and a generation step where we output NL explanation ($\explanation_q$) populated with the retrieved content in line with the template for the $et$. Hence, the synthesis task lends itself well to an application of the widely popular Retrieval-Augmented-Generation (RAG)~\cite{gao2023retrieval}, ~\cite{lewis2020retrieval} technique of prompting LLMs to output relevant answers. We use LlamaIndex~\cite{llamaindex}, particularly the PandasQueryEngine, to execute RAG on the explainer outputs using prompts that are enriched with the template for the identified explanation type $et$ and the user question, $uq$ to generate the final natural-language explanation $E$. We generate two explanations for each question (Listing \ref{lst:synthesis} in Appendix), one that summarizes the explainer outputs and the other that summarizes the subset of records that matched the filters in $uq$.

\subsection{System training setup}
Since we have used LLMs in our MetaExplainer framework, we trained, generated, and inferred on servers with multiple A100 GPUs. We conducted all our experiments between April - and June 2024 and, hence, used LLMs and methods that were state-of-the-art (SOTA) at that given time. In Sec. \ref {sec:discussion}, we discuss how system developers could introduce new methods to our framework for improved performance.

\subsection{User study setup} \label{sec:user-study-setup}
Since the MetaExplainer generates five multi-type explanations (Tab. \ref{tab:explanation_types}), each providing different insights to help end-users in their decision-making when using AI models, we wanted to understand the usefulness of our framework to potential end-users. Hence, we first carefully reviewed the various explanation evaluation metric dimensions proposed by Hoffman et al.~\cite{hoffman2018metrics} to analyze which ones best apply to our context, and we narrowed down the dimensions of evaluating the MetaExplainer on trust, satisfaction, and curiosity. We then designed three questionnaires - a per-explanation satisfaction questionnaire, overall trust and curiosity assessment, and user familiarity assessment questionnaire - to ask study participants to assess the MetaExplainer's performance and its explanations. Our study participants included a diverse cohort of $20$ university students and researchers working with various aspects of AI. We randomly selected a sample of $15$ question-explanation pairs for each participant, and we asked them to rank the explanations based on either Likert ratings or a $0\-2$ scale. The participants had no prior knowledge of the MetaExplainer or its workings; hence, their views reflect first-time system users. 
\section{Results} \label{sec:results}
\subsection{Datasets} \label{apt:datasets}
For the purpose of demonstrating the utility of the MetaExplainer in a high-precision use case such as healthcare, we wanted to select a comprehensive and simple dataset to be readily used in a single-variable classification task, e.g., whether a patient has Diabetes. We found the Pima Indians Diabetes Dataset~\cite{smith1988using}, a well-cited resource released and collected by the National Institute of Diabetes, Digestive and Kidney Diseases among high-risk Diabetes Mellitus ethnic Pima tribe women at least 21 years or older in New Mexico and Arizona regions in 1988. 
Recently, Chang et al.~\cite{chang2023pima} published their findings on which models work best with this dataset, and we used their findings to decide upon three ML classifiers (Logistic Regression - LR, Decision Tree - DT and Random Forest - RF). We find that the LR model has the best performance (F1 - 0.77) among the three models, and we have included these results in the Appendix (Sec. \ref{sec:appendix:diabetesmodels}). 

\subsection{Quantitative Results - Per-stage Evaluation} \label{sec:quant_results}
We evaluate the MetaExplainer at each stage using community-suggested metrics to assess methods used for similar tasks (i.e., classification and generation metrics for Decomposition, model explanation output evaluation metrics for Delegate, and retrieval metrics for Synthesis). In the interest of space, we present the parsing metrics for Decompose (Tab. \ref{tab:delegate_parsing}) and include the identification-based scores in the Appendix (Sec. \ref{sec:appendix:decompose}). 
As seen in Tab. \ref{tab:delegate_parsing}, we find that the fine-tuned Llama3 model has the highest F1-score on capturing the Likelihood field ($81.46\%$) in the $rq$, and for the machine-interpretation field that is used by the Delegate stage, the F1-score is $59.06\%$. In the Delegate stage, we evaluated the output of explainer methods based on quantitative metrics defined by the Explainable AI community for the modalities that these outputs are in~\cite{van2021evaluating}. 
As seen in Tab. \ref{tab:qual_perexplanationmetrics}, we find that the explainer methods produce faithful feature importances ($0.71$) and select a diverse set of samples ($340$). Finally, for the Synthesis stage, we evaluate the natural-language explanations by how relevant and faithful the explanations are to the provided context and user question. We find that the explanations are accurate and relevant to context (Tab. \ref{tab:synthesis-results}), indicating that the explanations provided are close to the results of the explainer outputs. 


\begin{table*}[hbt!] 
\centering
\caption{Performance metrics for text fields from Llama3 fine-tuned model (fine-tuned for 12 epochs on 170 question-reframed question pairs using LORA SFTTrainer) used in Decompose stage of the MetaExplainer.}
\label{tab:delegate_parsing}
\begin{tabular}{lccc}
\toprule
\textbf{Field} & \textbf{F1 (\%)} & \textbf{Precision (\%)} & \textbf{Recall (\%)} \\
\midrule
Machine Interpretation & 59.06 & 55.91 & 62.58 \\
Action                 & 57.48 & 50.00 & 67.60 \\
Likelihood             & \textbf{81.46} & \textbf{84.34} & \textbf{78.77} \\
\bottomrule
\end{tabular}
\end{table*}


\begin{table*}[hbt!] 
    \centering
     \caption{Summary of metrics for model explanation outputs from the Delegate stage of the MetaExplainer.}
       \label{tab:metrics_summary}
       \begin{adjustbox}{width=1\textwidth}
    \begin{tabular}{lrrr}
        \toprule
        \textbf{Metric} & \textbf{Mean values} & \textbf{Modality} & \textbf{Explanation Type + Explainer Method} \\
        \midrule
        Average rule length & 2.39 & Rules & Rationale, RuleXAI\\
        Fidelity & 0.31 & Rules & Rationale, RuleXAI  \\
        Non-representativeness & 0.026 & Samples & Case-based, Data and Counterfactual, Protodash and DiCE\\
        Diversity & 340.96 & Samples & Case-based, Data and Counterfactual, Protodash and DiCE\\
        Faithfulness & 0.71 & Features & Contrastive, SHAP  \\
        Monotonicity & 0.095 & Features & Contrastive, SHAP  \\
        \bottomrule
    \end{tabular}
    \end{adjustbox}
\end{table*}

\begin{table}[hbt!] 
    \centering
    \caption{Results of RAG metrics~\cite{es2023ragas} for $170$ natural-language explanations across $5$ explanation types, generated by the Synthesis stage of the MetaExplainer.}
    \label{tab:synthesis-results}
    \begin{tabular}{lr}
        \toprule
        \textbf{Metric} & \textbf{Value (\%))} \\
        \midrule
        Answer relevance & 66 \\
        Faithfulness & 25 \\
        \textbf{Context-utilization} & \textbf{67} \\
        \bottomrule
    \end{tabular}
\end{table}

\subsection{Qualitative Results - Ratings from user-study} \label{sec:qualt_results}

Through an analysis of responses to the three evaluation questionnaires filled out by our $20$ study participants (Sec. \ref{sec:user-study-setup}), we find that the majority of our study participants ($>90\%$) responded positively that our MetaExplainer helped them build trust in the AI and were able to satisfy their curiosity by our explanations (Tab. \ref{tab:likert_trust_curiosity}). Also, results from the per-explanation questionnaire demonstrate that users consistently responded positively across different explanation types (Tab \ref{tab:qual_perexplanationmetrics}). However, interestingly, only $67\%$ of the participants were positive in rating that the MetaExplainer satisfied their overall needs (Tab. \ref{tab:likert_trust_curiosity}), mainly they still have some doubts regarding how the MetaExplainer itself is working, thereby leading to some levels of hesitance in using the system for decision-making, indicating scope for improvement (Fig. \ref{fig:overall:deep-dive-users} in Appendix). 
We do not present results from the user familiarity questionnaire to uphold our study participants' identities; although the responses were anonymized, we want to be careful.

\begin{table}[hbt!]
\centering
\caption{Overall system satisfaction analysis (trust and curiosity) using Likert scale ($>=$Neutral).}
\label{tab:qual_perexplanationmetrics}

\begin{tabular}{p{6cm}p{1.5cm}}
\toprule
Evaluation Question & Percent Satisfied \\
\midrule
Overall positive respondents& 67.41 \\
\cmidrule{2-2}
I want to know more about what the AI did. & 93.33 \\
I want to know what the AI would have done if something had been different.  & 86.67 \\
I want to know why the AI did not make some other decision. & 86.67 \\
I want to understand what the AI will do next. & 80.00 \\
The outputs of the MetaExplainer are predictable. [Memorability] & 73.33 \\
I was surprised by the AI’s actions and want to know what I missed.  & 66.67 \\
The MetaExplainer can perform the task better than a novice human user. & 53.33 \\
\cmidrule{2-2}
I am confident in the MetaExplainer. I feel that it works well. [Error frequency] & 33.33 \\
I like using the system for decision-making. [Satisfication] & 33.33 \\
\bottomrule
\end{tabular}

\end{table}

\section{Discussion and Conclusion} \label{sec:discussion}
In this paper, we describe the MetaExplainer, a general-purpose framework to respond to user questions along multiple user-centered explanation types (Tab. \ref{tab:explanation_types}). The ability to provide explanations along multiple types helps tailor the explanations to the user's requirements~\cite{dey2022human},~\cite{liao2020questioning} and helps support explainers that can generate insights upon which these explanations depend. We implement the MetaExplainer as a modular three-stage framework (decompose-delegate-synthesis - Sec. \ref{sec:methods}), allowing us to run the different AI models separately needed to provide the eventual explanation (i.e., parsers in decompose, the explainers in delegate and natural-language generators) and have these models chain of each of their outputs. In this manner, we can also identify and use SOTA AI methods (e.g., LLama3 in decompose, AIX-360 explainers in delegate and RAG frameworks in Synthesis) demonstrated to be of use for the different tasks at each stage (Sec. \ref{sec:methods}). We can help system developers apply our MetaExplainer framework to different datasets and data modalities beyond tabular datasets and also swap in / add new explainer methods and LLMs at different stages. We make available the codebase as an open-source repository (linked in the abstract), facilitating easy adoption and implementation by the community. 



We have implemented several quantitative and qualitative evaluation strategies (Sec. \ref{sec:results}) to help end-users understand the MetaExplainer better and give system developers cues about the system. We are one of the few in the community (in comparison to TalkToModel's evaluation of the user utterances alone~\cite{slack2023explaining}) to implement quantitative evaluation strategies (Sec. \ref{sec:quant_results})) at each stage using community-accepted metrics for each model and task, improving traceability and error identification of the MetaExplainer. Also, our approach is a start towards developing quantitative evaluations of explanations~\cite{lakkaraju2022rethinking},~\cite{van2021evaluating} to help quantify and improve their impact and utility for upstream tasks~\cite{ghassemi2021false}. 

Our qualitative results show several trends, including differences in performance between explanation types (Tab. \ref{tab:qual_perexplanationmetrics} in Appendix), indicating value in individually investigating performance for each explanation type at each stage. Further, our qualitative results indicate that end-users are curious and eager to use our MetaExplainer (Tab. \ref{tab:likert_trust_curiosity}), which can provide explanations from different perspectives and several traces of detail, including outputs at each stage. However, they are not satisfied with the content (Tab. \ref{tab:qual_perexplanationmetrics} and \ref{tab:likert_trust_curiosity}), and would appreciate improved presentations. Several researchers adopt multiple user engagement sessions~\cite{wang2019designing},~\cite{slack2023explaining}, to improve the explanations AI systems can provide, which is a natural next step for us too. 

Finally, in implementing the MetaExplainer, we have leveraged the strengths of symbolic mappings, such as the EO~\cite{chari2023explanation}, which allows us to make minimal edits to a resource like EO, to support new explanation types/metrics/methods. This neuro-symbolic coupling is also an important step towards tractable development of explainer methods for different user needs, wherein both knowledge representation resources should be maintained for a growing body of literature and these can then be leveraged by frameworks providing explanations such as the MetaExplainer. Overall, the MetaExplainer is a step towards automatically supporting user-centered, multi-dataset and method explanations in response to user questions in different use cases.
\bibliography{references}

\appendix

\section{Motivation for MetaExplainer} \label{apd:motivation_for_metaexplainer}
Previous work in healthcare AI~\cite{gruen2021designing}~\cite{chari2023informing} has demonstrated the need and utility of explanations that are composed of different knowledge sources and explainer methods. As stated earlier, model explainer outputs on their own can be overwhelming for end-users to interpret, lack the grounding in the domain knowledge or context, and are often misleading on their own~\cite{ghassemi2021false}. Hence, end-users, specifically domain experts, prefer explanations that provide answers to a wide range of user questions such as the Why, Why not, What ifs, What cases, etc~\cite{liao2020questioning},  ~\cite{liao2022connecting}. Additionally, domain experts have a deep knowledge of the use cases but less so of the AI system, and either want to probe the decisions made by the system to understand its reasoning or learn more about the accuracy and trustworthiness of the decisions or globally learn about the data distributions the system operated on, or about global behavior. Hence, a framework such as the MetaExplainer that can provide answers in the form of diverse explanations for model decisions in various use cases is useful. The development of such a system introduces challenges of scalability in terms of minimal effort to load for new use cases, generalizability across a wide range of use cases, and interoperability with existing explainers. In our current implementation of the MetaExplainer (Fig.~\ref{fig:MetaExplainer-workflow}), we prioritize these attributes and develop a modular architecture that can be spun up with minimum human intervention, improved, and adapted at each stage.
\section{Explanation Types supported in MetaExplainer} \label{apt:explanation_types}
Listed are explanation types we currently support in the MetaExplainer along with methods that generate them (Tab. \ref{tab:qual_perexplanationmetrics}).
\begin{table}[hbt!]
\centering
\caption{Explanation types we currently support in the MetaExplainer}
\label{tab:explanation_types}

\begin{tabular}{|p{0.35\linewidth} | p{0.6\linewidth}|}
 \hline
Explanation Type &
  Definition \\  \hline
Case-based &
  Provides solutions that are based on actual prior cases that can be presented to the user to provide compelling support for the system’s conclusions and may involve analogical reasoning, relying on similarities between features of the case and of the current situation. \\ \hline
Contrastive &
  Answers the question “Why this output instead of that output,” making a contrast between the given output and the facts that led to it (inputs and other considerations), and an alternate output of interest and the foil (facts that would have led to it). \\  \hline
Counterfactual &
  Addresses the question of what solutions would have been obtained with a different set of inputs than those used. \\ \hline
Data &
  Focuses on what the data is and how it has been used in a particular decision, as well as what data and how it has been used to train and test the ML model. This type of explanation can help users understand the influence of data on decisions. \\ \hline
Rationale &
  About the “why” of an ML decision, and provides reasons that led to a decision, and is delivered in an accessible and understandable way, especially for lay users. \\ \hline
\end{tabular}%
\vspace{-0.5em}
\end{table}
\section{Methods}
\subsection{Decompose}
\begin{figure}[hbt!]
\centering
\includegraphics[width=1.0\linewidth]{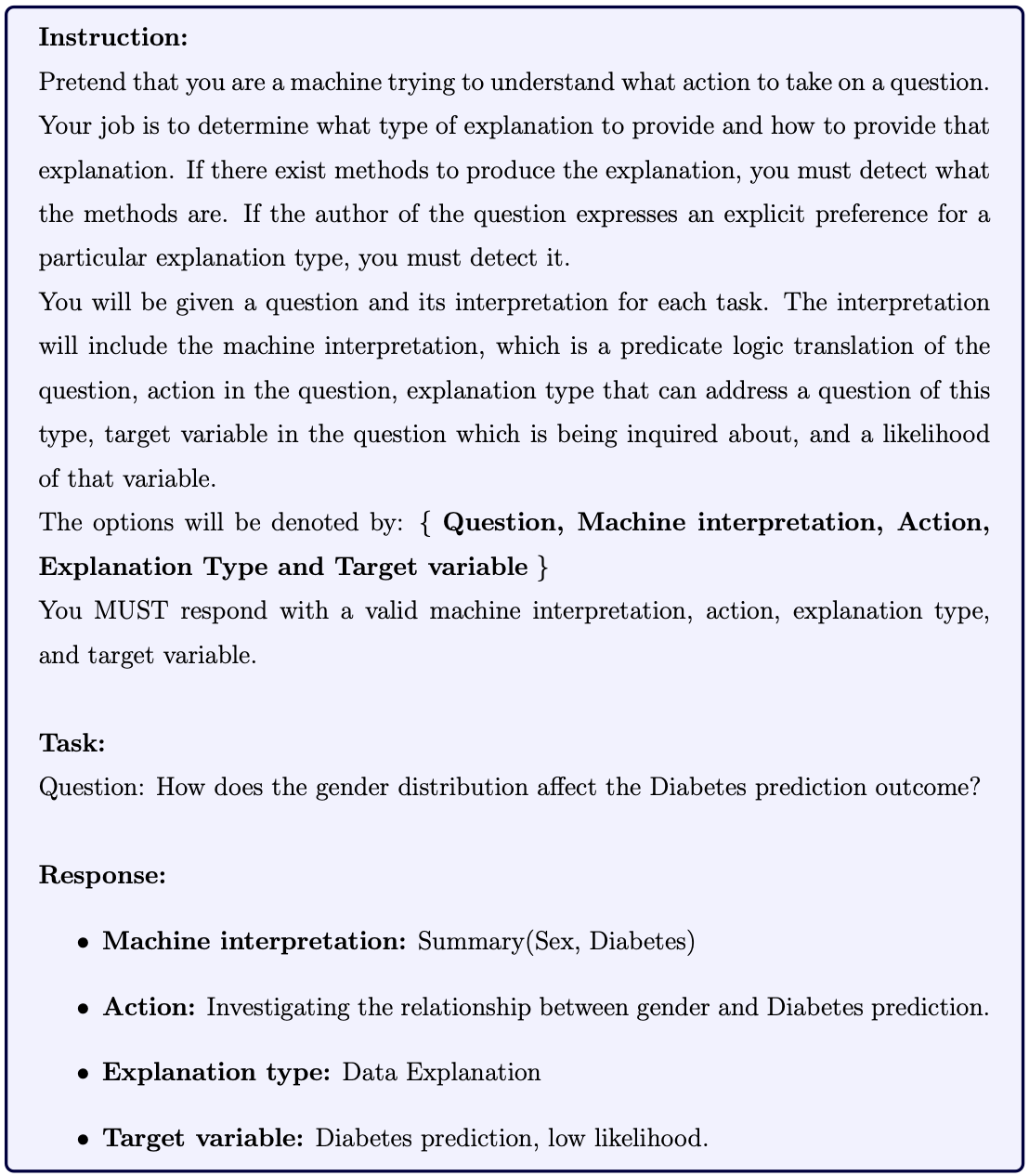}
\caption{Instruction prompt example used for fine-tuning Llama models to decompose question $uq$ into response, $rq$.}
\label{fig:prompt2}  
\end{figure}

The prompt used to identify the different parts of a user question for which explanations can be provided is as seen in Fig. \ref{fig:prompt2}.
\subsection{End-to-end example of user question being addressed by the MetaExplainer} \label{apt:end-end}
\begin{figure}[hbt!]
    \centering
\lstset{language=SPARQL, basicstyle=\ttfamily\fontsize{9}{10}\selectfont, xleftmargin=5mm, framexleftmargin=5mm, numbers=left, stepnumber=1, breaklines=true, breakatwhitespace=false, numbersep=5pt, tabsize=2, frame=lines}
\begin{lstlisting}
Question: How did the model justify predicting Diabetes for a 55-year-old male with a BMI of 18 and a Diabetes Pedigree Function of 0.25?

Explanation Type: Rationale Explanation

Machine Interpretation: 
    Predict(Diabetes, Age = 55, Sex = Male, BMI = 18, DPF = 0.25)

RuleXAI output:
    1. IF BMI = (<32.25, inf) THEN label = {1} 
    2. IF BMI (-inf, 37.05) THEN label = {1}
    
Explainer Method Metrics: 
    Average rule length - 1.4
    Fidelity: 0.4
\end{lstlisting}
\caption{Decompose and Delegate outputs from the MetaExplainer for an example question in a Diabetes prediction use case}
\label{lst:decompose_delegate}
\end{figure}

Below is an end-to-end example of outputs from each stage of our MetaExplainer framework. As seen in Listing \ref{lst:decompose_delegate}, a user question is translated to its predicate logic equivalent, and an explanation type of Rationale explanation is identified in the \textit{decompose} stage. The machine interpretation and explanation type fields are then passed to the \textit{delegate} stage, wherein the RuleXAI post-hoc explainer method is run that outputs rules that might have been used in the classification model, which have an average rule length of 1.4 and fidelity of 0.4. Finally, as seen in Listing \ref{lst:synthesis}, the rules and the original question are used by the \textit{synthesis} stage to provide natural-language summaries of the subset of data that matched the feature group restrictions in the question and the rules that were found to be responsible for the classification model predicting Diabetes for a 55-year old male with a BMI of 18 and Diabetes Pedigree Function of 0.25. In this way, we can see how each MetaExplainer stage builds on the previous stages' outputs and that a modular multi-stage framework such as ours helps capture restrictions in the question that otherwise might be missed by a pure LLM approach. 

\begin{figure}[hbt!]
    \centering
\lstset{language=SPARQL, basicstyle=\ttfamily\fontsize{9}{10}\selectfont, xleftmargin=5mm, framexleftmargin=5mm, numbers=left, stepnumber=1, breaklines=true, breakatwhitespace=false, numbersep=5pt, tabsize=2, frame=lines}
\begin{lstlisting}
Explanation of Matched subset:
    There are no full matches in the dataset based on the specified feature group. However, the dataset has been summarized, showing the descriptive statistics for the variables in the dataset. 
    - The mean age is 68.1 years.
    - The mean BMI is 29.49.
    - The mean Diabetes Pedigree Function (DPF) is 0.4868.
    - The outcome variable has a mean value of 0.3, indicating that the majority of the observations have an outcome of 0.

Explanation of explainer output:
    The rules derived for classification based on the features 
    'Age' and 'BMI' are as follows:
    1. IF BMI is less than or equal to 32.25, THEN label = 0
    2. IF BMI is greater than 37.05, THEN label = 1

These rules provide a clear guideline for classifying data based on BMI values, with different labels assigned depending on the BMI range. This type of explanation helps users understand the rationale behind the decision-making process in machine learning models, allowing them to assess the reasoning and make informed judgments.
\end{lstlisting}
\caption{Natural-language Explanations of Diabetes Prediction from Synthesis stage of MetaExplainer}
\label{lst:synthesis}
\end{figure}
\section{Results}
\subsection{Datasets} \label{sec:datasets}
For the purpose of demonstrating the utility of the MetaExplainer in a high-precision use case such as healthcare, we wanted to select a comprehensive and simple dataset to be readily used in a single-variable classification task, e.g., whether a patient has Diabetes. We found the Pima Indians Diabetes Dataset~\cite{smith1988using}, a well-cited resource released and collected by the National Institute of Diabetes, Digestive and Kidney Diseases among high-risk Diabetes Mellitus ethnic Pima tribe women at least 21 years or older in New Mexico and Arizona regions in 1988. 
The dataset could be considered small (768 records) in today's deep learning age, but it is still sufficient to apply simpler ML models such as logistic regression and decision tree classifiers. Recently, Chang et al.~\cite{chang2023pima} published their findings on which models work best with this dataset, and we used their findings to decide upon three ML classifiers (Logistic Regression - LR, Decision Tree - DT and Random Forest - RF). We find that the LR model has the best performance (F1 - 0.77) among the three models, and we have included these results here. The Pima Dataset comprises 768 rows and 9 columns - 8 features and 1 outcome variable. 
Although there were no direct missing column values in the dataset, like non-numeric values, there were zeroes for columns that were not collected or were missing for the patient instances. We applied data imputation techniques to fill in the median value of the column for zero values in the columns.

\subsection{Performance of fine-tuned Llama models in Decompose}
We evaluate the Decompose stage by how similar the reframed questions $rq$ generated by our fine-tuned LLMs are to the verified $rq$: $uq$ question pairs. We use F1 style metrics for the natural-language fields (Action, Machine Interpretation, and Likelihood) in $rq$ and report the classification accuracy for explanation types. Among the F1 metrics, we calculate the F1 by exact match~\cite{schneider2024evaluating} and Levenshtein distances. While Llama2 (see Tab. \ref{tab:confusion_matrix-llama2}) is slightly better at identifying explanation types than Llama3 (see Tab. \ref{tab:confusion_matrix-llama3}) for certain explanations (e.g., contrastive explanations), we choose Llama3 outputs over Llama2 since the parsing accuracy (see Tab. \ref{tab:results-decompose-llama3}) is far better.

The spread of human-verified questions generated by the GPT-3.5 Turbo model can be viewed from Tab. \ref{tab:explanation_counts_gpt}. We include a consoldiated file of the user question-reframed questions pairs across explanation types, used to fine-tune the LLama models in the supplementary material zip.

\begin{table}[ht!]
    \centering
       \caption{Explanation types of GPT-generated questions, $uq$ and their counts in the Decompose stage of the MetaExplainer.}
    \label{tab:explanation_counts_gpt}
    \begin{tabular}{lc}
        \toprule
        \textbf{Explanation Type} & \textbf{Count} \\
        \midrule
        \textbf{Data Explanation} & \textbf{80} \\
        \textbf{Case Based Explanation} & \textbf{60} \\
        Rationale Explanation & 50 \\
        Contextual Explanation & 35 \\
        Contrastive Explanation & 29 \\
        Counterfactual Explanation & 25 \\
        \bottomrule
    \end{tabular}
 
\end{table}

\begin{table*}[hbt!] 
\centering
\caption{Confusion matrix metrics for identification of different explanation types on $53$ sample test set, by Llama-3 fine-tuned model in the Decompose stage of the MetaExplainer.}
\label{tab:confusion_matrix-llama3}
\begin{tabular}{lcccc}
\toprule
\textbf{Explanation Type} & \textbf{Precision} & \textbf{Recall} & \textbf{F1-Score} & \textbf{Support} \\
\midrule
Contextual Explanation     & 0.50 & 0.67 & 0.57 & 6  \\
Data Explanation           & \textbf{0.80} & 0.62 & 0.70 & 13 \\
Contrastive Explanation    & 0.00 & 0.00 & 0.00 & 5  \\
Case Based Explanation     & 0.50 & 0.07 & 0.12 & 14 \\
Rationale Explanation      & 0.69 & \textbf{0.92} & \textbf{0.79} & 12 \\
Counterfactual Explanation & 0.50 & 1.00 & 0.67 & 3  \\
\midrule
micro avg                  & 0.60 & \textbf{0.51} & \textbf{0.55} & 53 \\
macro avg                  & 0.50 & 0.55 & 0.47 & 53 \\
weighted avg               & 0.57 & \textbf{0.51} & \textbf{0.48} & 53 \\
\bottomrule
\end{tabular}
\end{table*}

\begin{table*}[hbt!]
\centering
\caption{Confusion matrix metrics for identification of different explanation types on $53$ sample test set, by Llama-2 fine-tuned model in the Decompose stage of the MetaExplainer.}
 \label{tab:confusion_matrix-llama2}
\begin{tabular}{lcccc}
\hline
\textbf{Explanation Type} & \textbf{Precision} & \textbf{Recall} & \textbf{F1-Score} & \textbf{Support} \\ \hline
Contextual Explanation    & 0.00               & 0.00            & 0.00              & 6                \\ 
Data Explanation          & 0.67               & 0.46            & 0.55              & 13               \\ 
Contrastive Explanation   & 0.50               & 0.20            & 0.29              & 5                \\ 
Case Based Explanation    & \textbf{1.00 }              & 0.14            & 0.25              & 14               \\ 
Rationale Explanation     & \textbf{1.00 }              & 0.17            & 0.29              & 12               \\ 
Counterfactual Explanation& \textbf{1.00}               & \textbf{1.00 }           & \textbf{1.00 }             & 3                \\ 
\hline
\textbf{micro avg}        & 0.70               & 0.26            & 0.38              & 53               \\ 
\textbf{macro avg}        & 0.69               & 0.33            & 0.39              & 53               \\ 
\textbf{weighted avg}     & 0.76               & 0.26            & 0.35              & 53               \\ \hline
\end{tabular}

\end{table*}

\begin{table*}[hbt!] 
\centering
\caption{Performance metrics for text fields from the Llama2 fine-tuned model in Decompose stage of the MetaExplainer.} \label{sec:appendix:decompose}
\label{tab:results-decompose-llama2}
\begin{tabular}{lccc}
\toprule
\textbf{Field} & \textbf{F1 (\%)} & \textbf{Precision (\%)} & \textbf{Recall (\%)} \\
\midrule
\multicolumn{4}{l}{\textbf{F1 Exact Match scores on text fields}} \\
\midrule
Machine Interpretation & 46.34 & 61.29 & 37.25 \\
Action                 & 57.74 & 57.64 & 57.84 \\
Likelihood             & \textbf{50.14} & \textbf{59.35} & \textbf{43.40} \\
\midrule
\multicolumn{4}{l}{\textbf{F1 Levenshtein scores on text fields}} \\
\midrule
Machine Interpretation & 11.32 & 11.32 & 11.32 \\
Action                 & 8.25  & 9.09  & 7.55  \\
Likelihood             & 47.17 & 47.17 & 47.17 \\
\midrule
\multicolumn{4}{l}{\textbf{Exact match on text fields}} \\
\midrule
Machine Interpretation & \multicolumn{3}{c}{23.17} \\
Action                 & \multicolumn{3}{c}{28.87} \\
Likelihood             & \multicolumn{3}{c}{\textbf{25.07}} \\
\bottomrule
\end{tabular}
\end{table*}

\begin{table*}[hbt!] 
\centering
\caption{Performance metrics for text fields from Llama3 fine-tuned model used in Decompose stage of the MetaExplainer.}
\label{tab:results-decompose-llama3}
\begin{tabular}{lccc}
\toprule
\textbf{Field} & \textbf{F1 (\%)} & \textbf{Precision (\%)} & \textbf{Recall (\%)} \\
\midrule
\multicolumn{4}{l}{\textbf{F1 Exact Match scores on text fields}} \\
\midrule
Machine Interpretation & 59.06 & 55.91 & 62.58 \\
Action                 & 57.48 & 50.00 & 67.60 \\
Likelihood             & \textbf{81.46} & \textbf{84.34} & \textbf{78.77} \\
\midrule
\multicolumn{4}{l}{\textbf{F1 Levenshtein scores on text fields}} \\
\midrule
Machine Interpretation & 18.87 & 18.87 & 18.87 \\
Action                 & 19.23 & 19.61 & 18.87 \\
Likelihood             & \textbf{81.13} & \textbf{81.13} & \textbf{81.13} \\
\midrule
\multicolumn{4}{l}{\textbf{Exact match on text fields}} \\
\midrule
Machine Interpretation & \multicolumn{3}{c}{29.53} \\
Action                 & \multicolumn{3}{c}{28.74} \\
Likelihood             & \multicolumn{3}{c}{\textbf{40.73}} \\
\bottomrule
\end{tabular}
\end{table*}

\subsection{Best Performing Classification Models on Diabetes Dataset} \label{sec:appendix:diabetesmodels}
Here, in Tab. \ref{tab:model_performance_delegate}, we provide F1-scores on classification performance of various machine-learning models on classifying patients in the PIMA Indians Diabetes dataset. We used the best-performing Logistic Regression (LR) model ($77\%$) in our experiments. 

\begin{table*}[hbt!]
    \centering
       \caption{Model performance metrics on PIMA Indians Diabetes dataset.}
    \label{tab:model_performance_delegate}
    \begin{tabular}{lccccc}
        \toprule
        \textbf{Model} & \textbf{Precision} & \textbf{Recall} & \textbf{F1} & \textbf{Sensitivity} & \textbf{Specificity} \\
        \midrule
        Logistic Regression & \textbf{0.77} & \textbf{0.77} & \textbf{0.77} & 0.61 & \textbf{0.86} \\
        Decision Tree & 0.73 & 0.73 & 0.73 & \textbf{0.63} & 0.79 \\
        Random Forest & 0.75 & 0.75 & 0.75 & \textbf{0.63} & 0.82 \\
        \bottomrule
    \end{tabular}
\end{table*}

\subsection{Qualitative Results - Questionnaire files}
We include within the Github repository, questionnaires (\url{https://anonymous.4open.science/r/metaexplainer-C2BE/data/defaults/user_evaluations/}) we provided users with to rank explanations from the MetaExplainer. 

\subsection{User Study Additional Results}
\begin{table*}[hbt!]
\caption{MetaExplainer performance for satisfaction as rated by users.
 The first row shows the percentage of users with positive ($>=$Neutral (i.e., not unfavorable) on the Likert scale) assessment, and the other rows show a breakdown of Likert scores shown for each explanation type.}
\label{tab:likert_trust_curiosity}
\begin{tabular}{p{5cm}p{2cm}p{1.5cm}p{2cm}p{1cm}p{1.5cm}}
\toprule
Predicted Explanation Type & Case-Based & Contrastive  & Counterfactual  & Data  & Rationale  \\
\midrule
{\bf Overall Positive Respondents ($\%$)} & 93.33 & 90.48 & 91.30 & 92.59 & \textbf{95.45} \\
\cmidrule{1-1}
{\bf Average Likert Scores ( 1- 5)} & & & & & \\ 

I understand this AI system correctly due to the explanation. & 3.00 & 3.79 & 3.62 & \textbf{4.08} & 3.48 \\
The explanation is sufficiently complete. & 3.14 & 3.53 & 3.43 & \textbf{3.88} & 3.32 \\
This explanation has sufficient detail. & 3.43 & 3.79 & 3.57 & \textbf{4.08} & 3.54 \\
This explanation is useful to my goals. & 3.14 & 3.89 & 3.48 & \textbf{4.28} & 3.60 \\
\bottomrule
\end{tabular}
    
\end{table*}

\begin{figure}[hbt!]
\includegraphics[width=0.9\textwidth]{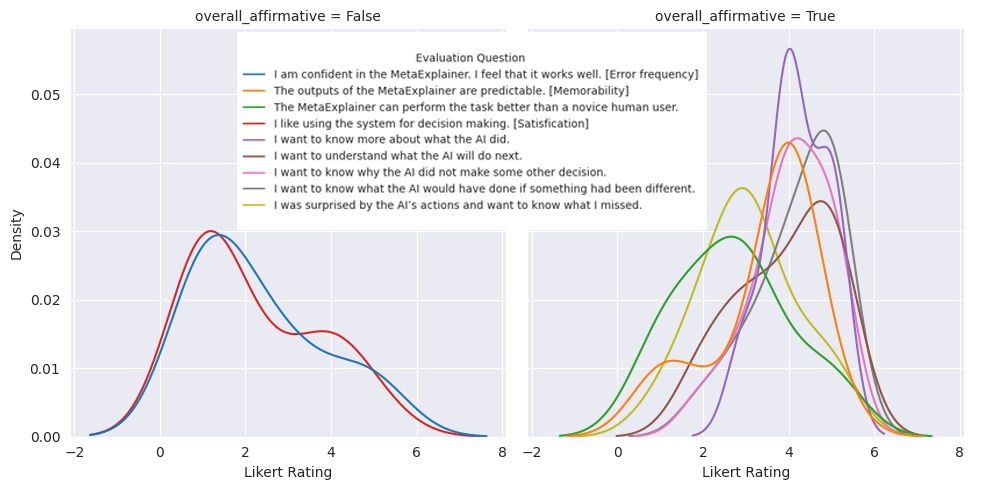}    
\caption{Distribution of Likert ratings for overall system performance. (left) shows the distribution of responses where users were not positive with regards to MetaExplainer, (right) shows that for the remaining questions, they were in general satisfied with MetaExplainer}
\label{fig:overall:deep-dive-users}
\end{figure}

\end{document}